\documentclass{aastex}
\begin{document}

\title{Distorted \ion{H}{1} Gas in the Widely Separated LIRG Arp 256}

\author{Jacqueline Chen\altaffilmark{1,2}}
\email{jchen@oddjob.uchicago.edu}
\author{K. Y. Lo\altaffilmark{1}}
\email{kyl@asiaa.sinica.edu.tw}
\author{Robert A. Gruendl\altaffilmark{3}}
\email{gruendl@astro.uiuc.edu}
\author{Miao-Ling Peng\altaffilmark{4}}
\and
\author{Yu Gao\altaffilmark{5}}
\email{gao@ipac.caltech.edu}
\altaffiltext{1}{Institute of Astronomy and Astrophysics (ASIAA), 
Academia Sinica, P.O. Box 23-141, Taipei 106, Taiwan}
\altaffiltext{2}{Department of Astronomy \& Astrophysics, University of 
Chicago, 5640 S. Ellis Ave., Chicago, IL 60637}
\altaffiltext{3}{Laboratory for Astronomical Imaging, Department of Astronomy,
University of Illinois, 1002 W. Green Street, Urbana, IL 61801}
\altaffiltext{4}{Department of Earth Sciences, National Taiwan Normal
University, Taipei, Taiwan}
\altaffiltext{5}{IPAC, MS 100-22, Caltech, 770 S. Wilson Ave., Pasadena, CA 
91125}

\begin{abstract}

We present new interferometric \ion{H}{1} and  CO ($J = 1 \rightarrow
0$)  observations of the luminous infrared source, Arp 256.  Arp 256
consists of two spiral galaxies in an early stage of merging, with a
projected nuclear separation  of 29 kpc ($54\arcsec$) and an infrared
luminosity of $2.0 \times 10^{11}~ L_{\sun}$.  Despite the large
separation of the galaxies' nuclei and mildly disrupted stellar
components, the \ion{H}{1} disks are found to be strongly disrupted,
and the southern galaxy in Arp 256 shows an elevated star formation
efficiency, which is consistent with a nuclear starburst.  Both of
these results run contrary to expectations, posing interesting
questions on the physical mechanisms involved in stimulating star
formation during an interaction.

\end{abstract}


\section{Introduction}

Luminous infrared galaxies (LIRGs) are galaxies with $L_{\rm IR}
\gtrsim 10^{11} L_{\sun}$ and in which most of the bolometric
luminosity is emitted in the infrared (see \citet{san96} for a
review).  In addition, they are often found in interacting or merging
systems -- with a frequency ranging from $\sim 10\%$ at
$L_{\rm{IR}}=10^{10.5}-10^{11} L_{\sun}$ to $\sim 100\%$ at
$L_{\rm{IR}}>10^{12} L_{\sun}$ -- and are  rich in molecular gas
\citep{san91,sol92,sol97}.   A majority of interacting systems,
however, are not LIRGs, so interactions are unlikely to be the sole
sufficient condition for explaining LIRGs \citep{kee85,bus86,ken87}.

Numerical simulations of interacting systems suggest that LIRGs might
be powered by nuclear starbursts.  Starbursts are triggered as tidal
interaction strips angular momentum from molecular gas in the galaxies
and gas clouds are drawn into the circumnuclear region during the late
stages of interaction \citep{mih96,bar96}.  This scenario
does not explain off-nuclear starbursts and gas concentrations
observed in some systems \citep{sta90,sar91,mir98}, although some
simulations do show enhanced extra-nuclear star formation (e.g.,
Fig. 13, \citet{mih96}).  In addition, numerical simulations do not
necessarily reflect real star formation processes, since star
formation is modeled, by the Schmidt Law, as a function of gas
density only.  Previous observational work on LIRGs have found nuclear
starbursts and gas concentrations  consistent with the numerical
results, but these studies generally focused on late-stage mergers and
ultraluminous infrared galaxies (ULIRGs), which lack the initial
condition information that would be useful in comparisons to
simulations \citep{kee85,san96}.

Few studies have explored mergers of gas-rich galaxies over the entire
merger sequence.  \citet{hib96} studied a range of phases in the
merger sequence in the \ion{H}{1}, \ion{H}{2}, and  $R$-bands for a
small sample of galaxies.  They suggest that \ion{H}{1} disks should
become more disrupted as the merger progresses, from $\sim 60 \%$ of
the $M_{\rm HI}$ being found in the main disks and bulges  of each
system at an early stage to near 0\%, late in the merger sequence.  In
addition, \citet{gao99} have shown that star formation efficiency (SFE
= $L_{\rm IR}/M_{\rm H_{2}}$) increases as separation between
interacting galaxy pairs decreases.  In general, these results, as
well as other observational results and numerical simulations suggest
that the starburst phase of interacting galaxies occurs at a late
stage in the interaction and is confined to the central kpc.

Arp 256 is composed of two galaxies separated by $\sim54\arcsec$.  Arp
256 N  (north) is a SB(s)c galaxy, while Arp 256 S (south) is a SB(s)b
galaxy.  At a redshift of $z=0.02715, ~cz=8140$ km $\rm s^{-1}$, their
distance has been calculated at 98.1 Mpc ($H_{0}$=75 km $\rm s^{-1}$
$\rm Mpc^{-1}$) with a projected nuclear separation of 29 kpc
\citep{con90,sur93}.  The IR luminosity based on IRAS observations is
$2.0 \times 10^{11}~L_{\sun}$, but neither galaxy is known to have AGN
activity and both show \ion{H}{2} region-like spectra \citep{vei95}.
The relatively large separation between the galaxies in Arp 256
suggests that the system is in an early phase of merging and, thus, is
of interest as an observational example of a LIRG in the early stages
of the formation process.

\section{Observations and Data Reduction}

Table \ref{ta1} summarizes the details of all the
observations.  \ion{H}{1} 21 cm observations of Arp 256 were conducted
with the Very Large Array (VLA) in the 'C' configuration on 1998
December 14.  The time  on source was nearly 7 hours, and the
half-power beamwidth of $\sim 32 \arcmin$ was centered between the two
galaxies and fully covered all the emission with a velocity bandwidth
of 63 channels of 21.75 km $\rm s^{-1}$ each.  The observing frequency
was 1.3826 GHz, corresponding to 8159 km $\rm s^{-1}$ in the optical
definition of velocity with respect to the Local Standard of Rest. The
AIPS software package was used for the calibration and data reduction.
The data was carefully flagged and two calibration sources, quasars
3C48 (IAU name: 0137+331) and 0022+002, were used to derive the
antenna-based gain and the bandpass solutions.  The continuum was
estimated from the line-free channels (channels 1-24 and 45-63,
velocity ranges 8833 - 8333 and 7876 - 7485 km $\rm s^{-1}$) and
subtracted in the $u-v$ plane with UVLIN to result in a data cube with
only \ion{H}{1} emission.  \ion{H}{1} emission is clearly detected in
the velocity range 8290 - 7963 km $\rm s^{-1}$ (channels 26-41), a
velocity linewidth of 370 km $\rm s^{-1}$.  Two image maps were made
from this data cube.  One resultant map uses a synthesized beam of $19
\farcs 2 \times 14 \farcs 8$ and a ROBUST = 0 weight function to
achieve 0.38 mJy $\rm beam^{-1}$ rms noise level.  An additional,
``naturally'' weighted map, with synthesized beam of $24 \farcs 8
\times 18 \farcs 7$, was made and used to construct a velocity contour
map (see Figure \ref{mom1}).

A radio continuum image was extracted from the the emission-free
channels (channels 3-22, 43-61, velocity ranges 8790 - 8377 and 7920 -
7528 km $\rm s^{-1}$) of the \ion{H}{1} dataset.  A ROBUST=-5,
``uniform'' weight function was used in order to maximize the
resolution of the image, resulting in a $17 \farcs 7 \times 13 \farcs
5$ synthesized beam and 0.61 mJy $\rm beam^{-1}$ rms noise level.

CO $(J = 1 \rightarrow 0)$ in Arp 256 was observed in 1998 May and
November at the Berkeley-Illinois-Maryland Association (BIMA)
10-element millimeter wave array in the 'C' configuration.  Arp 256
was observed for 28 minutes, alternated with 6.5 minute observations
of the phase and amplitude gain calibrator 0006-063 for a total time
on source of 9 hours.  Mars and Uranus were observed as flux
calibrators.  The pointing was centered between the galaxies and the
half-power beamwidth of $\sim 2\arcmin$ covered the central 64 kpc.
The observing frequency was 112.2204 GHz, corresponding to 8149 km
$\rm s^{-1}$ in the optical convention.  In addition, 78 channels of
19.9 km $\rm s^{-1}$ velocity width provided full frequency coverage.
Line emission was found in channels 27-48 -- velocity range 7890 -
8308 km $\rm s^{-1}$ -- for a velocity linewidth of 440 km $\rm
s^{-1}$.  Calibration and data reduction were done with the MIRIAD
software package.  The resultant map uses a synthesized beam of $11
\farcs 5  \times 4 \farcs 9$ and natural weighting to achieve a 0.038
Jy $\rm beam^{-1}$ rms noise level.
 
\section{Results}

\subsection{Spatial Distribution of the \ion{H}{1} Emission}  

In Figure~\ref{over1}, the \ion{H}{1} column density map is overlaid
on a $J$-band image of Arp 256 from a mosaic of 52 exposures of 60
seconds each from observations made with the NIRIM camera
\citep{mei99} on the Mount Laguna Observatory 1-m telescope.  The
\ion{H}{1} distribution shows a spiral pattern, but, when compared to
the near-infrared image, the \ion{H}{1} emission peaks do not
correspond to the optical emission peaks of the galaxies.  This
suggests that the original \ion{H}{1} disks associated with the two
galaxies have been severely disrupted despite the large separation
between the two galaxies.  The strong disruption of the \ion{H}{1}
disks may indicate that the galaxies have passed the period of first
closest approach.
 
The integrated flux measured in the VLA map is 6.6 Jy km $\rm s^{-1}$, 
which is essentially the same as the single-dish flux of 6.01 Jy km 
$\rm s^{-1}$, measured by \citet{bus87}.  From the integrated flux, 
the total \ion{H}{1} mass can be estimated by
\begin{displaymath}
   M_{\rm H I} (M_{\sun}) = 2.36 \times 10^5 D^2 \int Sdv,
\end{displaymath}
where $D$ is distance in Mpc, and $\int Sdv$ is \ion{H}{1} integrated
flux in Jy km $\rm s^{-1}$ \citep{rob75}.  The  value of $\int Sdv$ is 
obtained by summing the flux in the channel maps.  The resultant total 
\ion{H}{1} mass is $1.5 \times 10^{10} M_{\sun}$.

The \ion{H}{1} channel maps and the position-velocity map show at
least four distinct gas components (see
Figures~\ref{hichan},\ref{sv1}):  the tail of \ion{H}{1} emission
extending from the  northern arm of Arp 256 N; the central peak of
emission between the two galaxies; a small amount of gas that overlaps
the optical emission of Arp 256 S on the sky; and the emission peak to
the south-west of Arp 256 S.  These components have $\sim$20, 30, 5,
and 15\% of the integrated flux, and therefore of the total \ion{H}{1}
mass, respectively.

\subsection{Spatial Distribution of the CO Emission}

Figure~\ref{over1} also shows the integrated CO intensity map overlaid
on the optical image.  In contrast to the \ion{H}{1} distribution, the
CO emission associated with Arp 256 S appears unresolved, as well as
undisturbed, and well-confined within the circumnuclear regions of the
galaxy.  No CO emission is detected in the nuclear regions of Arp 256
N, although there is a marginal, $3\sigma$ detection in the vicinity
of the southern major spiral arm.  The absence of detected emission in
Arp 256 N is unexpected for a late-type  spiral (Sc) galaxy.  In
general, early-type spirals -- Sab and Sb -- have a $M_{\rm
H_{2}}/M_{\rm HI}$ ratio around 2 and spiral galaxies later than Sc
have a ratio smaller than 1 \citep{you89}, but the original \ion{H}{1}
disks of Arp 256 have been disrupted, so this comparison is no longer
possible.

We estimate the molecular gas mass from the integrated CO(1-0) flux 
using the empirical relation \citep{sco87}:
\begin{displaymath}
        M_{\rm H_2}(M_{\sun}) = 1.18 \times 10^4 D^2 \int Sdv,
\end{displaymath}
where $D$ is distance in Mpc and $\int Sdv$ is CO integrated flux in
Jy km $\rm s^ {-1}$.   Here the Galactic conversion factor $X \equiv
N(\rm H_2)$/$I_{\rm CO}=3.0 \times 1 0^{20}$  $\rm cm^{-2}$ \rm (K km
$\rm s^{-1})^{-1}$ (see \citet{you91} for a review) is assumed for
convenient  comparison with previous results.   We note that this
conversion factor may cause  an overestimation of the  molecular gas
mass if the galaxy is undergoing starbursts (e.g.,
\citet{mal88,bry99}).  The integrated flux of the CO emission
associated with the southern nucleus is 42 Jy km  $\rm s^{-1}$,
resulting in an estimated $\rm H_{2}$ mass for the southern nucleus of
$4.8 \times 10^{9} M_{\sun}$.  The $1\sigma$ upper limits to the
integrated flux and the $M_{\rm H_{2}}$ for the northern source are 29
Jy km $\rm s^{-1}$ and 3.3 $\times 10^{9} M_{\sun}$, respectively, employing 
the same linewidth as Arp256 S.

\subsection{Spatial Distribution of the Radio Continuum}

A 21 cm radio continuum map was extracted from line-free channels of
the VLA \ion{H}{1} observations and displayed in Figure~\ref{over1}.
A strong, unresolved radio source overlaps the bulge of Arp 256 S.
Weaker continuum emission is detected in the direction of Arp 256 N,
with a peak over the northern major spiral arm and a weaker peak over
the southern major spiral arm (see Figure~\ref{radioclose}), with the
possibility that the radio emission forms a spiral to match  the
spiral arms of the galaxy.  Future observations should include a
higher resolution radio map in order  to follow up on this unusual
structure.

Measurements of the 20 cm radio continuum flux imply that the radio
continuum is emitted almost entirely from the circumnuclear region of
Arp 256 S -- inside a radius of 2 kpc.  \citet{con98} find an
integrated flux density of 43.2 mJy at a  resolution of $45\arcsec$,
while the VLA Faint Images of the Radio Sky at Twenty-Centimeters
(FIRST) Survey finds 34.82  mJy within a  $\sim4\arcsec$ beam
\citep{whi97}.  This corroborates the high level of activity in the
southern source as seen in our map and  implies the presence of a star
forming region or an AGN in Arp 256 S.  A closer inspection of the
southern galaxy may detect a slight offset between the peak in
emission of the radio continuum and the peak in emission of the CO
disk.  While the radio continuum emission appears to  be centered
directly on the $J$-band image peak, the CO emission appears slightly
to the north-east.  This offset, while intriguing, is similar to the
pixel size for both the CO map and the $J$-band map and has to be
verified by further observations.

\subsection{Kinematics}

Figure~\ref{mom1} shows isovelocity contours from the  first moment of
the \ion{H}{1} data cube.  The velocity field appears to have the
characteristics of two ``spider diagrams'' centered at the nuclei of
the two galaxies.  However, as the spectral lines are not single
peaked and the velocity gradients do not follow the major axes of the
optical galaxies -- as would have been expected for an isolated,
rotating, inclined disk -- the ``spider diagram'' features are likely
incidental.   The systemic velocities of the two galaxies appear very
similar, $\sim$8150 km $\rm s^{-1}$, measured via the optical
definition with respect to the Local Standard of Rest.  It is possible
that Arp 256 N has a slightly greater recessional velocity, 8175
compared to 8125 km $\rm s^{-1}$ for Arp 256 S.  In comparison,
heliocentric optical redshifts for the two galaxies show the northern
nucleus at 8193 km $\rm s^{-1}$ and the southern nucleus at 8112 km
$\rm s^{-1}$.  In addition, the CO map for Arp 256 S shows a velocity
width of 300 km $\rm s^{-1}$, centered around 8130 km $\rm s^{-1}$,
calculated via the optical definition.  The \ion{H}{1} gas appears to
``rotate'' about Arp 256 N, from south-south-west to north-east
between 8246 and 8050 km $\rm s^{-1}$, whereas the \ion{H}{1} around
Arp 256 S ``rotates'' from near the galaxy to west-south-west between
8159 and 8007 km $\rm s^{-1}$, asymmetric in distribution.

These rotational signatures may be misleading, since, in searching for
signatures of rotational motion in smaller slices of the
spatial-velocity diagram, it becomes apparent that very little gas can
actually be attributed to any particular galaxy.  For example,
Figures~\ref{nsv} and \ref{ssv} are spatial velocity diagrams for the
two galaxies of Arp 256 separately, summing emission perpendicular to
the approximate east-west alignment of the major axes of both galaxies.
They clearly show that despite the large amount of \ion{H}{1} gas in
the system, very little of the original \ion{H}{1} disks around Arp
256 N and S remain.  The declinations which correspond to  the optical
bulges of the galaxies appear at approximately the same velocity, 8115
km $\rm s^{-1}$.  In addition, the major peaks of emission (offset on
the sky from the galaxies) do not exhibit normal rotational motion and
appear essentially at one  velocity.  At lower velocities, the
northernmost peak of emission appears at 8070 km $\rm s^{-1}$, while
the southernmost peak is at 8030 km $\rm s^{-1}$.  At higher
velocities, the strongest peak of emission, between the two galaxies,
appears at two distinct velocities, 8160 and 8250 km $\rm s^{-1}$.

Any attempt to synthesize the kinematic data into an orbital geometry
for the system can be only partly successful, as we remain burdened
with too many degrees of freedom.  Given the lack of strong tidal
tails and the strong disruption of the \ion{H}{1} disks, it is likely
that at least one of the  spin vectors of the galaxies is retrograde
with respect to the orbital  angular momentum vector and a
prograde-prograde configuration is unlikely \citep{mih96}.   The
velocity contours show that the direction of the gradient in the two
galaxies is in opposite directions along an east-west axis.  However,
as the galaxies appear to be at least somewhat face-on with respect to
the plane of the sky -- Arp 256 S is  measured at an inclination of
$48\degr$ \citep{bot82,mar91} -- even this is inconclusive in
determining whether the galaxies' spin axes are aligned or in opposite
directions.

\section{Discussion}

Given the level of activity in the southern galaxy, it is appropriate
to ask whether the source is a starburst or an AGN.  The answer,
however, is difficult to determine definitely.  A simple way to
distinguish between a nuclear starburst and an AGN would be to resolve
the center of the galaxy;  a nuclear starburst would be confined to
$\lesssim$ 1 kpc of the center, while an AGN would be confined to  a
much smaller radius.  In addition, an offset of the radio continuum
emission from the gas that might feed star formation or from starlight
could distinguish a starburst from an AGN.  Current radio continuum
images of Arp 256 S are unable to resolve the center of the galaxy and
distinguish between a nuclear starburst or an AGN.  They resolve at
best to only 1-2 kpc, and the previously noted offset of the CO disk
is inconclusive, given the resolution of our images.  Starbursts
exhibit a strong far-infrared / radio  continuum correlation (see
\citet{con91}), which also proves inconclusive in the case of Arp 256,
falling within an acceptable range for a starburst, but not ruling out
an AGN \citep{lia00}.  However, previously published data of line
ratios taken from optical spectra distinguish between a starburst and
an AGN.  \citet{vei95} find, via optical spectra, that both of the
galaxies in Arp 256 are \ion{H}{2} region-like (not powered by an AGN)
and neither LINER nor Seyfert-like.  In addition, \citet{bus90} find
significant  H$\alpha$ emission in the southern nucleus, supporting
those results.

The star formation efficiency (SFE) of the galaxies can be estimated
by the  ratio $L_{\rm IR}(8-1000 \mu{\rm m})/M_{\rm H_{2}}$.  Given a
$L_{\rm IR}$ from \citet{car88,car90} of $0.87 \times
10^{11}~L_{\sun}$ for the northern source and  of $1.17 \times
10^{11}~L_{\sun}$ for the southern source, the SFE of Arp 256~S is
24.4 $L_{\sun}/M_{\sun}$, far higher than the typical SFE of giant
molecular clouds in the Milky Way \citep[$\sim 4
L_{\sun}/M_{\sun}$]{sco89} and within the  20-100 $L_{\sun}/M_{\sun}$
range for nuclear starburst galaxies \citep{sco91}.  The $L_{\rm IR}$
given by \citet{car88,car90}, however, is likely to be an
underestimate as it is based upon an extrapolation from near-IR
measurements and assumptions about the far-infared ratios.  On the
other hand, from the radio continuum map, it is clear that $\sim$75\%
of the emission is from the southern source.  The total IR luminosity
detected by IRAS is $2.0 \times 10^{11}~L_{\sun}$.  If we assume that
Arp 256 S emits $\sim$75\%, then we estimate a similar SFE for Arp 256
S of $\sim 30~L_{\sun}/M_{\sun}$.

It seems likely, then, that there is a starburst in the nuclear
regions of Arp 256 S.  The infrared luminosity and the molecular gas
mass give an estimate of how long it will take for the molecular gas
to run out.  If we assume the infrared luminosity is entirely due to
dust heated by young stars in recent star formation, the star
formation rate (SFR) is directly proportional to the infrared
luminosity, with a relation of SFR ($M_{\sun}$ $\rm yr^{-1}$) $\sim 1
\times 10^{-10}~L_{\rm IR}(L_{\sun}$) for stars with masses greater
than 2 $M_{\sun}$  \citep{gal86}.  For Arp 256, the estimated star
formation rate would be 11.7 $M_{\sun}~\rm yr^{-1}$.  This rate
essentially measures gas depletion, so given the inferred molecular
gas mass,  the molecular gas will be exhausted in $\sim 4 \times
10^{8}$ yr.  This timescale is inconsistent with the expectation of a
nuclear starburst at a late stage of merging for galaxies with central
bulges.  Given that Arp 256 is currently in an early-stage of merging
associated with the period of first close approach \citep{mih96}, the
starburst in Arp 256 S will be exhausted before the late stages of
merging.

In comparison to other studies, other discrepancies also arise.
\citet{gao99}, in a study of SFE in systems at various stages of
interaction, note a correlation between SFE and interaction time --
the longer the interaction time, the greater the SFE -- which supports
the standard model of starbursts at late stages of mergers.  In
comparison to that study, Arp 256 has  the highest SFE for its
apparent separation (the measure of interaction time used).  The
highly disrupted \ion{H}{1} disk and the low percentages of flux
associated with the optical disks of each galaxy also run contrary to
the results of \citet{hib96}, who find an  anti-correlation between
the stage of merger and the central \ion{H}{1} mass.  Other
observations, however, find cases of early-stage mergers with possible
starbursts (e.g., \citet{wan01}).

The lack of detected CO would imply that Arp 256 N has little star
formation activity.  \citet{vei95}, however, identify the galaxy as
\ion{H}{2} region-like.  In addition, the H$\alpha$-band images of Arp
256 N of \citet{bus90}, clearly show a peak of emission in the
northern arm but not in its nucleus, nor in the southern arm -- where
we had a marginal detection of CO and an upper limit of $M_{\rm H_{2}}
\lesssim 3.3 \times 10^{9} M_{\sun}$.  Our radio continuum map peaks
off-nucleus and may possibly overlap the H$\alpha$ peak in the
northern arm and is, therefore, consistent with the results of
\citet{vei95}.

\section{Conclusions}

From our new observations of \ion{H}{1} and CO in Arp 256 we see
evidence contradicting the standard model of LIRGs.  Despite
the large separation between the galaxies and no morphological evidence 
of interaction that would accompany an intermediate or late stage
merger, Arp 256 displays the highly disrupted \ion{H}{1} disks and
increased SFE commonly associated with a late stage of merging.  The 
interaction history of Arp 256 that has led it to its current set of
characteristics, however, are unavailable with current evidence and
models, and more study is required of this system and others like it.

\acknowledgements

We thank S.-W. Lee for assisting in the acquisition of data and
D.-C. Kim and D.B. Sanders for the use of the $R$-band image of Arp
256.  We are grateful to John Hibbard, the referee, for critical
comments which were very helpful in improving the paper.  We would
also like to thank W.-H. Wang for his invaluable comments and
suggestions.  The research has made use of the NASA/IPAC Extragalactic
Database (NED) which is operated by the Jet Propulsion Laboratory,
California Institute of Technology, under contract with the National
Aeronautics and Space Administration.  R. A. Gruendl acknowledges
support from the Laboratory of Astronomical Imaging, which is funded
by the NSF grant AST 99-81363 and by the University of Illinois.
M.-L. Peng started the VLA \ion{H}{1} data reduction while a summer
student at the ASIAA.  Research at the ASIAA is supported by the 
Academia Sinica in Taipei and the National Science Council of Taiwan.

\clearpage

\figcaption[Chen.fig1.ps]{Zero moment \ion{H}{1}, CO, and radio
continuum maps in contours over an optical image.  The optical image
is a $J$-band image  from the Mount Laguna Observatory (MLO) 1-m
telescope.  The \ion{H}{1} emission is marked by the green contours
and beamsize indicator.  Contour levels are 1, 2, 3, $\ldots$ , 12
times 0.04 Jy $\rm beam^{-1}$ km $\rm s^{-1}$, which corresponds to a
column density of $\rm 1.6 \times 10^{20} ~cm^{-2}$.  The blue
contours  mark the radio continuum image, in levels of  50, 150, 250,
500, 800, 1100, 1500, 1800, and 2100 times  0.61 mJy $\rm beam^{-1}$
$(1\sigma$ = 1.54 K).  The CO emission is given in red contours, also
with levels of 3, 5, 7, and 9  times 3.510 Jy $\rm beam^{-1}$ km
$s^{-1}$, which corresponds to  a molecular hydrogen column density of
$\rm 1.7 \times 10^{21} ~cm^{-2}$.
\label{over1}}

\figcaption[Chen.fig2.ps]{ROBUST = 0, uniform weighted \ion{H}{1}
channel maps, with $\sigma$ =  8.3 mJy $\rm beam^{-1}$ km $\rm s^{-1}$
levels overlaid on the MLO $J$-band image.  Contours are 2, 4, 6,
$\ldots$ , 14 times $\sigma$.  24-43 of 63 channels are plotted.  Each
frame is labelled in the top left corner with the line-of-sight
velocity in km $\rm s^{-1}$.
\label{hichan}}

\figcaption[Chen.fig3.ps]{\ion{H}{1} position-velocity $(p-v)$ diagram
of the summed emission perpendicuar to the north-south axis, both in
contours and grey-scale.  Contour levels are at 10\% of the peak at
51.5 mJy $\rm beam^{-1}$.  Four emission features are labelled:  the
tail of HI emission extending from the northern arm (NT), the central
peak between the two galaxies (CP), the small amount of gas that
overlaps the optical emission of Arp 256 S on the sky (OR), and the
emission peak to the southwest of Arp 256 S (SW). \label{sv1}}

\figcaption[Chen.fig4.ps]{Close-up of Figure \ref{over1}, with only
radio continuum contours overlaid on a $R$-band image of Arp 256 N
provided by D.B. Sanders and D.-C. Kim.  The $R$-band image consists
of one 300 second exposure from the Palomar 60-cm telescope.  Levels
are 40, 75, 110, 140, 170, 205, 250, and 280 times 0.61 mJy $\rm
beam^{-1}$ $(1\sigma$ = 1.54 K).  \label{radioclose}}

\figcaption[Chen.fig5.ps]{\ion{H}{1} radial velocity contour and
grey-scale map, derived from the naturally weighted map of Arp 256.
Contours are plotted every 25 km $\rm s^{-1}$, the central contours
falling at 8150 and 8175 km $\rm s^{-1}$.  \label{mom1}}

\figcaption[Chen.fig6.ps]{Spatial velocity diagram along the east-west
axis of Arp 256 N of emission summed from $\delta_{2000} =
-10 \degr 22 \arcmin 00 \arcsec$ to $\delta_{2000} = -10 \degr 21
\arcmin 00 \arcsec$, approximately perpendicularly to the major axis of
the galaxy.  Levels are 1, 2, 3, 4, 5 times 6.0 Jy $\rm beam^{-1}$.  
The nucleus of the optical galaxy falls at
$\alpha_{2000} = 00^{\rm h}18^{\rm m}50\fs0$ and 8115 km $\rm
s^{-1}$. \label{nsv}}

\figcaption[Chen.fig7.ps]{Spatial velocity diagram along the east-west
axis of Arp 256 S of emission summed from $\delta_{2000} =
-10 \degr 23 \arcmin 12 \arcsec$ to $\delta_{2000} = -10 \degr 22
\arcmin 20 \arcsec$, approximately perpendicularly to the major axis of
the galaxy.  Levels are 1, 2, 3, \ldots , 7 times 5.6 Jy $\rm beam^{-1}$.  
The nucleus of the optical galaxy
falls at $\alpha_{2000} = 00^{\rm h}18^{\rm m}50\fs7$ and 8115 km $\rm
s^{-1}$. \label{ssv}}

\clearpage
\begin{deluxetable}{lrr}
\tablecaption{Observing Parameters \label{ta1}}
\tablehead{\colhead{}  &
\colhead{21 cm Emission}  &
\colhead{CO ($J = 1 \rightarrow 0$)}}
\startdata
Telescope &  VLA & BIMA \\
Observation Date &  12/98  &  5/98, 11/98 \\
Velocity Center (km $\rm s^{-1}$): & & \\ 
~~~~~$cz$ & 7980 & 7933 \\
~~~~~Optical Convention & 8159 (w.r.t. LSR) & 8149 \\
Time on Source (hrs) & 7.0 &  9.1 \\
Number of Channels & 63 & 78  \\
Channel Separation (km $\rm s^{-1}$) & 21.75 & 19.9  \\
Line Velocity Range (km $\rm s^{-1}$) & 8290-7963 & 7890-8308 \\
Integrated Flux (Jy km $\rm s^{-1}$) & 6.6 & 29, 42\tablenotemark{a}\\
Affiliated Mass ($10^{9}~M_{\sun}$) & 15\tablenotemark{b} & 
3.3, 4.8\tablenotemark{c} \\
\enddata
\tablenotetext{a}{Integrated flux for Arp 256 N and Arp 256 S, 
respectively, where the value for the northern galaxy is the 1$\sigma$ 
upper limit.}
\tablenotetext{b}{Total \ion{H}{1} mass, $M_{HI}$.}
\tablenotetext{c}{Molecular gass mass, $M_{H_{2}}$, for Arp 256 N and 
Arp 256 S, respectively.}
\end{deluxetable}

\end{document}